\documentclass{article}
\usepackage{epsfig}
\usepackage[centertags]{amsmath}
\usepackage{graphicx}
\usepackage{cite}
\usepackage{url}

\begin{document}
\title{Time-of-Flight Signatures of Two Bosons in a Double Well}
\author{Analabha Roy and L.E. Reichl \\
Center for Complex Quantum Systems\\
and\\
Department of Physics\\
The University of Texas at Austin, Austin, Texas 78712\\}
\date{\today }
\maketitle
\begin{abstract}
We present analytical and numerical treatments for evaluating the time-of-flight momentum distribution for the stationary states of a two-boson system trapped in a quartic double-well potential, paying particular attention to the Tonks and noninteracting regimes. We find that the time-of-flight distributions can serve as a valuable tool in profiling the states of this system, and provide the tof plots for low-energy excitations using a nonadaptive finite element method that is more efficient than traditional finite difference methods.
\end{abstract}
\section{Introduction}
The study of ultracold boson systems in Optical Dipole Traps (ODT) such as Bose-Einstein condensates, have received considerable attention in academia over the last decade of the 20$^{th}$ century~\cite{weiman}~\cite{weiman:cornell}~\cite{ketterle}~\cite{ketterle2}. More recently, experiments involving number squeezed states of trapped alkali atoms have yielded promise~\cite{raizen}. New techniques, such as  quantum tweezing ~\cite{diener} and quantum many-body culling~\cite{raizen}, are being developed that can create mesoscopic two-boson systems out of ultracold atoms in optical traps. Theoretical studies demonstrate the possibility of number state generation by atomic culling as well, where a BEC is number-squeezed by "culling" atoms from a trapped condensate down to a sub-poissonian regime, making the number uncertainty small enough to be ignored~\cite{raizen}. Such a two-boson system can be subjected to a micrometer-scale double well by various means, ranging from small volume optical traps~\cite{raizen}, to atom chips~\cite{doublewell:chip}~\cite{doublewell:chip:nature}. An optical lattice of such double-wells can also be generated by two counter-propagating lasers of linearly polarized light with a known angle between their planes of polarization, and a transverse magnetic field to mix the two potentials~\cite{Deutsch:Jessen}. If the on-site lattice depth is sufficiently deep then the tunneling between the sites can be neglected. Furthermore, if they are loaded homogeneously from a cold-atom system confined in an optical dipole trap by atom culling~\cite{raizen}, each double well system can be treated in isolation exactly as depicted in~\cite{mypaper}. More recently (2009), number squeezing and subpoissonian distribution of atoms in each site in an optical lattice have been reported by Itah et al~\cite{technion:oplattice-culling}. The weak nature of the interactions of such atoms means that such cold atom systems are useful tools for quantum information processing~\cite{Monroe:Nature}, as well as in studying quantum entanglement~\cite{Jaksch:Zoller}. More recently, the study of the dynamics of quantum control in such systems have been performed as well~\cite{mypaper}. The momentum distributions of two-bosons systems have been evaluated numerically by Murphy and McCann using a finite difference method~\cite{McCann:doublewell}. We present the numerical evaluation of the time of flight momentum distributions of the lower energy stationary states of this system using a nonadaptive finite element method that is conputationally faster and more accurate.

In the following sections, we evaluate the time of flight (tof) signatures of these wavefunction. Section~\ref{sec:2} details how the double well system was diagonalized and the eigenfunctions obtained. Section~\ref{sec:3 } discusses the nature of the time-of-flight signatures of the different states, and numerical results are shown in section~\ref{sec:4 }. Concluding remarks are made in the final section.

\section{The Eigensystem: Strongly Interacting and Single Particle Regimes} 
\label{sec:2}
Our system consists of  two alkali metal bosons confined to a double-well optical potential.  The effective interaction between the bosons, in three dimensions, is obtained in the long wavelength approximation to be 
\begin{equation}
u^{3d}({\bf x}_1-{\bf x}_2)= \frac{4\pi\hbar^2 a_s}{m} \delta({\bf x}_1-{\bf x}_2),
\end{equation}
where ${\hbar}$ is Planck's constant, $a_s$ is the s-wave scattering length and ${\bf  x}_i = \left(  x_i,y_i,z_i \right) $ is the displacement of the $i$th particle~\cite{metcalf:vanderstraten}~\cite{pethick:bec}. The system can be confined in two spatial (radial) directions so that the essential dynamics occurs in the $x$ - direction by the use of  anisotropic magnetic traps with high aspect ratio~\cite{olshanii:1d}~\cite{petrov:1d}. In that case, the other 2 dimensions can be integrated out~\cite{mypaper}~\cite{olshanii:1d}, yielding an effective 1-dimensional interaction

\begin{equation}
u(x_1-x_2) = 4a_s \omega_s \hbar \delta(x_1-x_2)
\end{equation}

\begin{figure} 
\hspace*{-0.3in}
\ \psfig{file=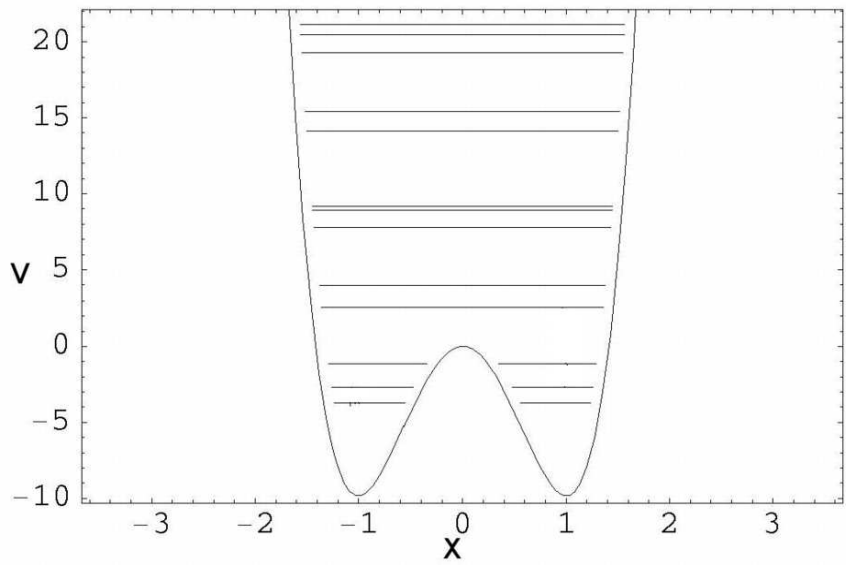,height=3.0in,width=5.0in}
\caption{Plot of the double-well potential experienced by each boson for the single particle regime. The energy levels, $E_1=-6.42262$, $E_2=-5.68883$ and $E_4=0.640055$  of the interacting two-boson system (interaction strength  $U_0=-1.0$) are also sketched. Here, $V_0=7.2912229$.}
\label{fig:doublewell_case02}
\end{figure}

We will consider the case of  two identical bosons  confined to a quartic double well potential. We get the total Hamiltonian for the system to be
\begin{equation}
H =p^2_1+p^2_2+V_0 (-2 x_1^2+ x_1^4) +V_0 (-2 x_2^2+ x_2^4)+U_0 \delta(x_1-x_2) .
\label{eq:hamscale }
\end{equation}

where $p_i$ is the momentum of the $i$th particle ($i$= 1,2), $x_i$ is the position of the $i$th particle along the x-axis,  and $V_0$ determines the depth of the double well potential. We have used dimensionless expressions for all the degrees of freedom, as well as the system parameters, by introducing a characteristic length scale $L_u$. Thus, the actual Hamiltonian $H'$ relates to the dimensionless Hamiltonian $H$ as  $H=\frac{H'}{E_u}$, where $E_u=\frac{\hbar^2}{2mL^2_u}$. Similarly,  $U_0=\frac{4a_s \omega_s \hbar}{E_u}$ and the time scales as $t=\frac{t'}{T_u}$ where $T_u=\frac{2mL^2_u}{\hbar}$. Fig.~\ref{fig:doublewell_case02} shows a plot of the quartic double well $V(x)=V_0 (-2 x^2+ x^4)$ for  well depth $V_0=7.2912229$.

The numerical diagonalization of the Hamiltonian in Eq.~\ref{eq:hamscale } is facilitated by a nonadaptive finite element method using the analytically obtained matrix elements of the Hamiltonian in a finite wave train basis of size $L=3.5$ (in units of $L_u$),
\begin{equation}
{\langle}x_1,x_2\vert n_1,n_2{\rangle} ^{(s)}=\frac{1}{\sqrt{2(1+\delta_{n_1,n_2})}} 
[{\langle}x_1|n_1\rangle{\langle}x_2|n_2\rangle +{\langle}x_1|n_2\rangle{\langle}x_2|n_1\rangle ].
\label{eq:symm}
\end{equation}
Here,
\begin{equation}
\langle x|n\rangle=\frac{1}{\sqrt{L}} \sin{\biggl[}{\frac{n\pi}{2}(\frac{x}{L}-1){\biggr]}}
\label{eq:pboxfn}
\end{equation}
within the range $-L\leq x \leq L$ and vanishes outside.

\begin{figure}
\hspace*{-0.5in}
\ \psfig{file=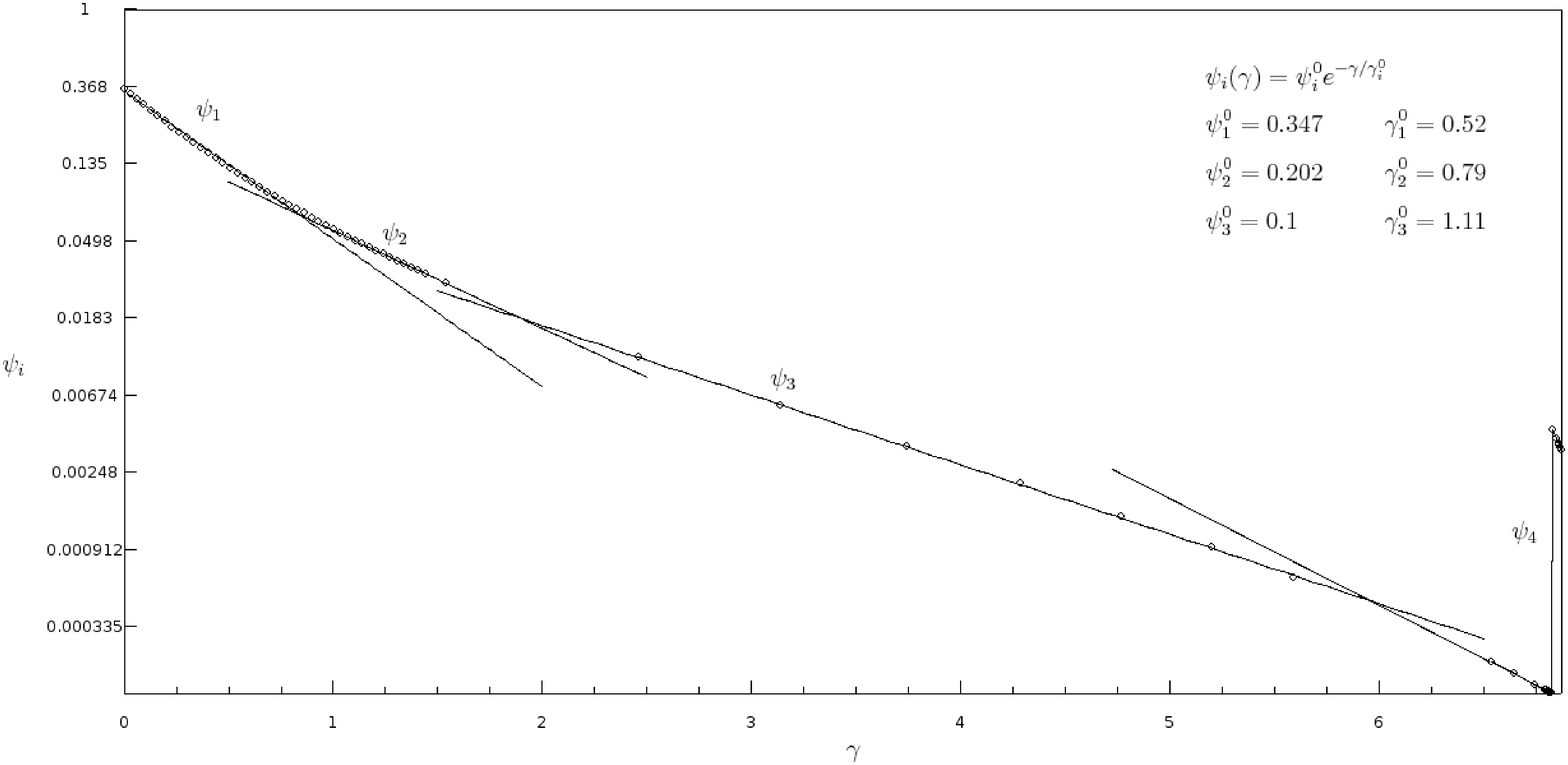,height=3.8in,width=6.0in}
\caption{Semi-logarithmic plot of the one dimensional probability density $p_i/{\delta l}$ of two particles being within a rectangular strip of arbitrarily small width $\delta l$ along the line $x_1=x_2$. $p_i/{\delta l}$ is plotted as a function of the strongly interacting parameter $\gamma$ for a constant $V_0=4.0$. The decay rate of the probability $p_i$ changes sharply at 4 regions, labeled by the index $i$. The data points (indicated by circles) have been fitted to exponential decay rates at each region (indicated by lines).The legend provides the numerically fitted values of the decay rates $\gamma^0_i$. Note the discontinuous spike at $\gamma\sim 7$.}
\label{fig:tonksparam }
\end{figure}

We will investigate the tof distributions in two regimes of the $\left(V_0,U_0\right)$ parameter space of the double well system. Here, $V_0$ is the well depth, and $U_0$ the amplitude of the point contact pseudopotential in 1-dimension.
The first regime, henceforth referred to as the 'strongly interacting regime' will consist of a very strongly repulsive system and a moderate well depth. We define the 'strongly interacting factor' for this system, $\gamma$, as 
\begin{equation}
\gamma \equiv \frac{U_0}{E}.
\end{equation}
Here, $E$, the energy of the state, is a measure of the ability of the bosons to tunnel across from one well to another. When $\gamma \rightarrow \infty$, we reach the strongly interacting regime where the interaction completely dominates the system~\cite{tonks:gas}. Figure~\ref{fig:tonksparam } shows the evolution of the ground state of the system as $\gamma$ is increased. The order parameter being plotted as a function of $\gamma$ for the ground state is $p_i/{\delta l}$, where 
\begin{equation}
p_i \equiv \delta l \int dx| \langle x , x | E_1 \rangle |^2.
\end{equation}

Here, $i$ is an index distinguishing different regimes of interest in the $\gamma$-space. Also, $p_i$ is the total probability that the two particles will be together within a rectangular strip along the line $x_1=x_2$ and arbitrarily small width $\delta l$. As expected, it vanishes for large values of $\gamma$.

In this strongly interacting regime, the two particles have no probability of occupying the same position simultaneously. Thus, they act in a way that is similar to a Tonks gas~\cite{tonks:gas}.  The transition to this regime is not consistent, however. We note four distinct ranges of $\gamma$ for which the decay rates of 
$p_i/{\delta l}$  are different. In the first three ranges, $p_i$ seems to be decaying exponentially ie $\left( p_i/{\delta l}\right)=\left( p^0_i/{\delta l} \right) e^{-(\gamma/\gamma^0_i)}$ for $i=1,2,3$. The data points have been fitted to exponents by the use of numerical nonlinear least-squares algorithms.
The decay rate, characterized by $ \gamma^0_i $, decreases sharply at $\gamma\sim 1, 2$ and $6$. Near $\gamma\sim 7$, there is a sharp increase in $p_i$ after which it continues to decrease. If we neglect the probability if it falls below $1/e$ of the maximum, then the 'strongly interacting regime' is achieved beyond $\gamma\sim 0.4$. In our case, we have chosen a $\gamma$ of $5.20142$ for our strongly interacting regime, placing the system in region $3$ of Fig~\ref{fig:tonksparam }. The value of $\left(V_0,U_0\right)$ chosen is $\left(4.0, 40.0\right)$.

\begin{figure}
\ \psfig{file=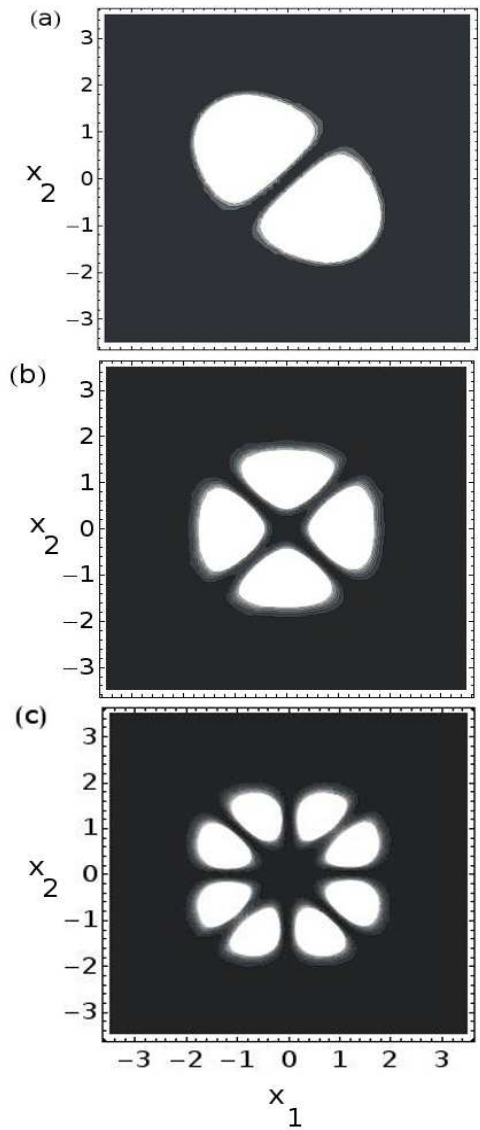,height=6.0in,width=3.0in}
\caption{Plots of energy eigenfunctions for the two interacting bosons in a double well potential in the strongly interacting regime. Figures (a) through (c) are contour plots of the probability density $|\langle x_1,x_2|E_1\rangle|^2$ ,  $|\langle x_1,x_2|E_2\rangle|^2$  and $|\langle x_1,x_2|E_4\rangle|^2$ respectively. The probabilities are plotted as functions of $x_1$ and $x_2$. All units for all figures are dimensionless}
\label{fig:wavefunctions_tonks }
\end{figure}

\begin{figure}
\ \psfig{file=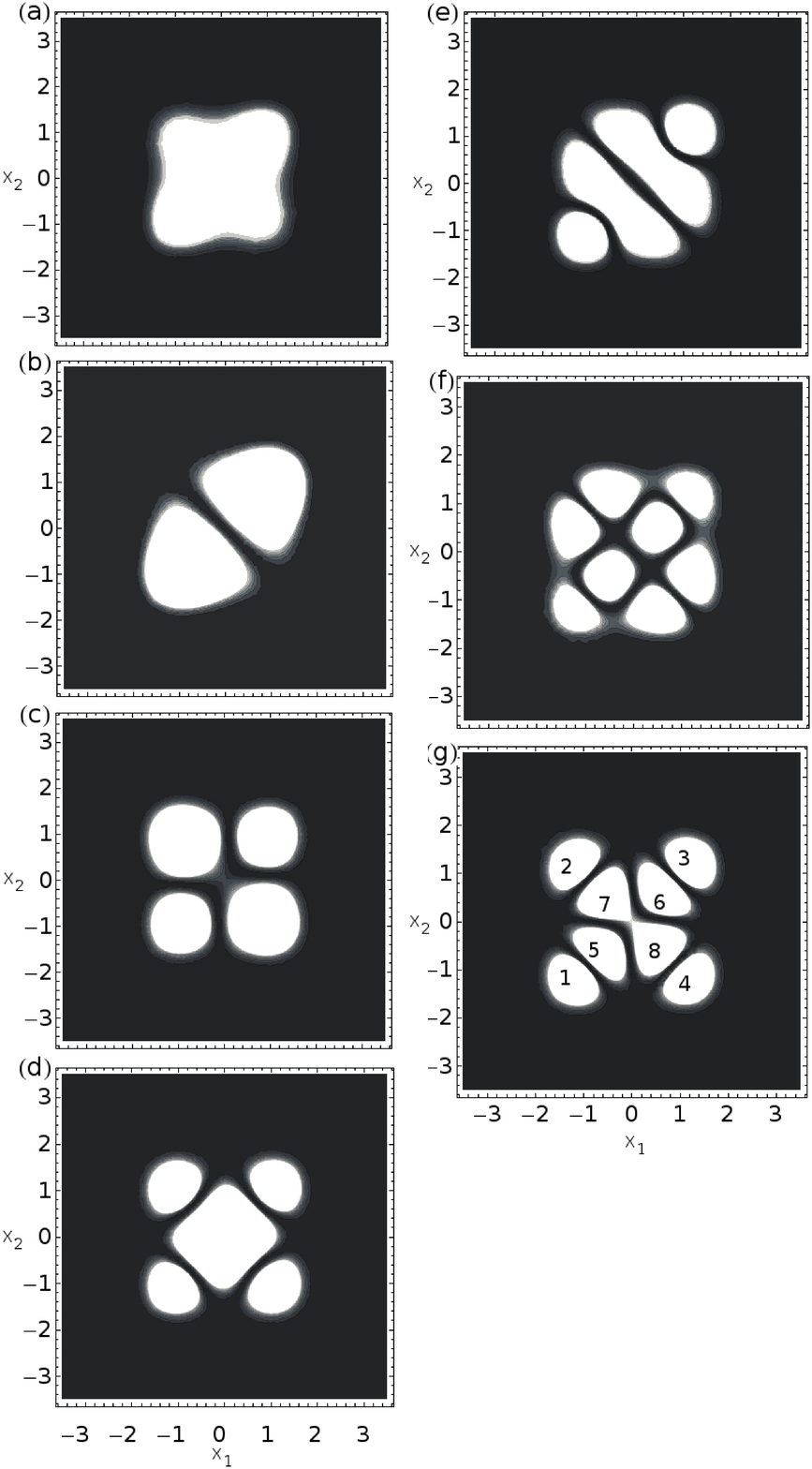,height=6.0in,width=4.0in}
\caption{Plots of energy eigenfunctions for the two interacting bosons in a double well potential in the single particle regime. All units are dimensionless. Figures (a) through (f) are contour plots of the probability density $|\langle x_1,x_2|E_1\rangle|^2$ through $|\langle x_1,x_2|E_6\rangle|^2$ respectively. Figure (g) is a contour plot of the probability density
$|\langle x_1,x_2|E_7\rangle|^2$.The peaks in the probability are numbered. The probabilities are plotted as functions of $x_1$ and $x_2$. All units for all figures are dimensionless}
\label{fig:wavefunctions }
\end{figure}

The second regime, henceforth referred to as the 'single particle regime', will consist of a weakly attractive system and the well-depth as seen in~\cite{mypaper}. Thus, the parameter values chosen are $\left(7.2912229, -1.0\right)$.The probability distributions of the ground state  $|E_1\rangle$, as well as the excited states  $|E_2\rangle$ and  $|E_4\rangle$, given by Eqn~\ref{eq:hamscale }  are shown in Figs~\ref{fig:wavefunctions_tonks }.a through \ref{fig:wavefunctions_tonks }.c for the strongly interacting regime. Note that, as expected, there is virtually no probability that $x_1=x_2$.
The probability distributions of the first seven quantum energy states of the system in the single particle regime are shown in Figs.~\ref{fig:wavefunctions }.a through  \ref{fig:wavefunctions }.g. Note the plots of the ground state, $|E_1\rangle$, third excited state, $|E_4\rangle$,  and sixth excited state $|E_7\rangle$. 

\section{ Time of Flight Images} 
\label{sec:3 }
The normalized first order correlation function of a single double well is a measurement of the atomic density $n(x)$. Such correlations can be measured by the time-of-flight (TOF) technique in which the trapped atoms are released sufficiently quickly  that the diabatic approximation in quantum mechanics can be applied. The atoms then expand ballistically until they reach a detection plate. If the plate is far enough from the double well system that the far-field approximation can be used, then the Green's Function for the system can be simplified and the time translation reduced to a simple Fourier Transform.  The 'detector plane' coordinates are denoted by unprimed variables $\left[ x_1,x_2,t \right]$ and the double -well coordinates by primed variables $\left[ x'_1,x'_2,t \right]$ for a 2-particle system after all the external fields and traps have been diabatically switched off. The interactions between the atoms while in flight can be rendered negligible by tuning a homogeneous magnetic field close to the Feshbac resonance that adds an attractive amplitude to the normally repulsive point contact pseudopotential~\cite{feshbach:resonance}~\cite{pethick:bec}. The system then evolves ballistically in free space. 

The Green's Function or Propagator $G({\bf x},t;{\bf x}',t')$ is defined by
\begin{equation}
\Psi ({\bf x},t) = \int d^2x \mbox{ } G({\bf x},t;{\bf x}',t') \Psi ({\bf x}',t'),
\label{eq:greensfunction:appendix}
\end{equation}
where ${\bf x} = \left[ x_1, x_2 \right]$, and $\Psi({\bf x},t)$ is the wavefunction, with similar expressions for the primed coordinates. For free space, the relevant 1-dimensional Schr\"odinger equation for 2-particles is
\begin{eqnarray}
\left[ H-i\frac{\partial}{\partial t} \right] \Psi({\bf x},t) =0, \nonumber \\
H=-\left[ \frac{\partial^2}{\partial x^2_1} + \frac{\partial^2}{\partial x^2_2} \right].
\end{eqnarray}
 Thus, the Green's function~\cite{sakurai} will be the solution to
 \begin{equation}
 \left[ H-i\frac{\partial}{\partial t} \right] G({\bf x},t;{\bf x}',t') = \delta ({\bf x}-{\bf x}') \delta (t-t').
 \label{eq:greens:appendix}
 \end{equation}
The solution to Eqn~\ref{eq:greens:appendix} in free space is
\begin{equation}
G({\bf x},t;{\bf x}',t') = \frac{\sqrt{-i}}{L} \exp{\left[ i\pi | \frac{{\bf x}-{\bf x}'}{L}|^2 \right]}.
\label{eq:greensfn:appendix}
\end{equation}
Here, the ballistic de-Broglie equation,
\begin{equation}
L^2=4 \pi \tau,
\label{eq:debroglie }
\end{equation}
 provides the relation between the detector-system separation $L$ and the time-of-flight $\tau=\left(t-t'\right)$.

Now, consider such a two particle system localized at a site $j$. The wavefunction is localized about ${\bf x}'_j = \left[ x'_j, x'_j \right]$ and can be written in the form $\Psi({\bf x}'-{\bf x}'_j)$. We now use Eqns~\ref{eq:greensfn:appendix} and~\ref{eq:debroglie } on Eqn~\ref{eq:greensfunction:appendix}, and apply the shift theorem for Fourier transforms~\cite{goodman}~\cite{Grondalski:etal} to get
\begin{equation}
\Psi(x,\tau) = \sqrt{\frac{-i}{4\pi\tau}}\exp{\left[ i\frac{1}{2\tau}\left( \frac{|{\bf x}|^2}{2} + {\bf x}\bullet {\bf x}'_j\right)\right]} F\left[ \Psi({\bf x}')\right]_{{\bf u}=\frac{{\bf x}}{4\pi\tau}},
\label{eq:TOF }
\end{equation}
where the primed coordinates refer to the double well system, the unprimed coordinates refer to the detector, and $ F\left[ \Psi({\bf x}')\right]_{\bf u}$ is the Fourier transform
\begin{equation}
F\left[ \Psi({\bf x}')\right]_{\bf u} \equiv \frac{1}{2 \pi} \int d^2x' \Psi({\bf x}') e^{i{\bf u}\bullet{\bf x}'}.
\end{equation}
In the equation above,  ${\bf u} = \left[ u_1, u_2 \right]$ is the momentum space vector. For a large collection of such systems, each in the desired pure state, the measured TOF is simply the probability obtained from Eqn~\ref{eq:TOF } times the number of such double wells $N$ (which we shall subsequently drop off as an appropriately adjusted overall normalization).
\begin{equation}
n({\bf x})= N \frac{1}{4\pi\tau} | F\left[ \Psi({\bf x}') \right]_{{\bf u}=\frac{{\bf x}}{4\pi\tau}} |^2.
\label{eq:TOFmeas }
\end{equation}

In the next section, Eqn~\ref{eq:TOFmeas } will be evaluated numerically for two bosons in a double well, and the density functional 
\begin{equation}
n(x) = \int dx'' n(x,x'')
\end{equation}
will be calculated in order to provide the 1-dimensional tof distribution.

\section{ Time-of-Flight: Numerical Plots}
\label{sec:4 }
This section will detail the procedure for obtaining numerical plots of the tof distributions of the eigenstates of the double well system. The two boson problem in a double well is diagonalized as detailed in section~\ref{sec:2} and~\cite{mypaper}. Thus, the eigenfunctions are obtained as linear superpositions of the eigenfunctions of two bosons in a box of appropriately chosen length $L$, ie
\begin{eqnarray}
{\langle}x_1,x_2\vert n_1,n_2{\rangle} ^{(s)}=\frac{1}{\sqrt{2(1+\delta_{n_1,n_2})}} 
[{\langle}x_1|n_1\rangle{\langle}x_2|n_2\rangle +{\langle}x_1|n_2\rangle{\langle}x_2|n_1\rangle ], \nonumber \\
\langle x|n\rangle=\frac{1}{\sqrt{L}} \sin{\biggl[}{\frac{n\pi}{2}(\frac{x}{L}-1){\biggr]}},
\label{eq:pboxfn:appendix}
\end{eqnarray}
 if $|x_i|<L$, and $0$ otherwise. Thus, the final solution to an eigenfunction $|E_n\rangle$ of the double well will be a linear superposition of the  'finite wave train'  functions defined above, ie
\begin{equation}
\langle x_1,x_2|E_n\rangle =  \sum_{\left[n_1,n_2\right]=\left[1,1\right]}^{\left[ N,N \right]} C^{\left[n_1,n_2 \right]}_{E_j} {\langle}x_1,x_2\vert n_1,n_2{\rangle} ^{(s)},
\end{equation}
where the $C^{\left[n_1,n_2 \right]}_{E_j}$ are obtained numerically using the nonadaptive finite element method. This result can then be substituted into Eqn~\ref{eq:TOFmeas } to get
\begin{equation}
n(x_1,x_2) = N \frac{1}{4\pi\tau} |\sum_{\left[n_1,n_2\right]=\left[1,1\right]}^{\left[ N,N \right]} C^{\left[n_1,n_2 \right]}_{E_j} F\left[ {\langle}x'_1,x'_2\vert n_1,n_2{\rangle} ^{(s)} \right]_{\left[u_1,u_2\right]=\frac{\left[ x_1,x_2 \right]}{4\pi\tau}}|^2.
\label{eq:density:appendix}
\end{equation}
Using the linearity of Fourier Transforms and Eqn~\ref{eq:pboxfn:appendix}, we get
\begin{multline}
F\left[ {\langle}x'_1,x'_2\vert n_1,n_2{\rangle} ^{(s)} \right]_{\bf u}=\frac{1}{\sqrt{2(1+\delta_{n_1,n_2})}}\\
\left( F[{\langle}x_1|n_1\rangle]_{u_1} F[{\langle}x_2|n_2\rangle]_{u_2} +F[{\langle}x_1|n_2\rangle]_{u_1} F[{\langle}x_2|n_1\rangle]_{u_2}  \right).
\label{eq:ft:appendix}
\end{multline}
The Fourier transform of the finite wave train ($\langle x|n\rangle$ in Eqn~\ref{eq:pboxfn:appendix}) can be calculated using Gaussian integrations~\cite{arfken} to yield
\begin{equation}
F[\langle x' | n \rangle]_u = -I\sqrt{\frac{L}{2\pi }}\left \{\frac{\sin{\left(uL+\frac{n\pi }{2}\right)}}{\left(uL+\frac{n\pi}{2}\right)} e^{-i \left(n\pi /2\right)}- \frac{\sin{\left(uL-\frac{n\pi }{2}\right)}}{\left(uL-\frac{n\pi}{2}\right)}e^{i \left(n\pi /2\right)} \right \},
\label{eq:pboxft }
\end{equation}
where $u$ is the momentum space vector. Thus, by plugging Eqn~\ref{eq:pboxft } into Eqn~\ref{eq:ft:appendix}, and that into Eqn~\ref{eq:density:appendix}, the tof distribution $n(x_1,x_2)$ can be obtained, the final density distribution is the density functional average of this result viz.
\begin{equation}
n(x)=\int dx' n(x, x').
\end{equation}
Thus, a numerical expression for Eqn~\ref{eq:TOFmeas } was obtained for two degrees of freedom $x_1$ and $x_2$, and the density functional $n(x)$ determined by integrating out one of the coordinates by adaptive Gauss-Kronrod quadrature. 

\begin{figure}
\ \psfig{file=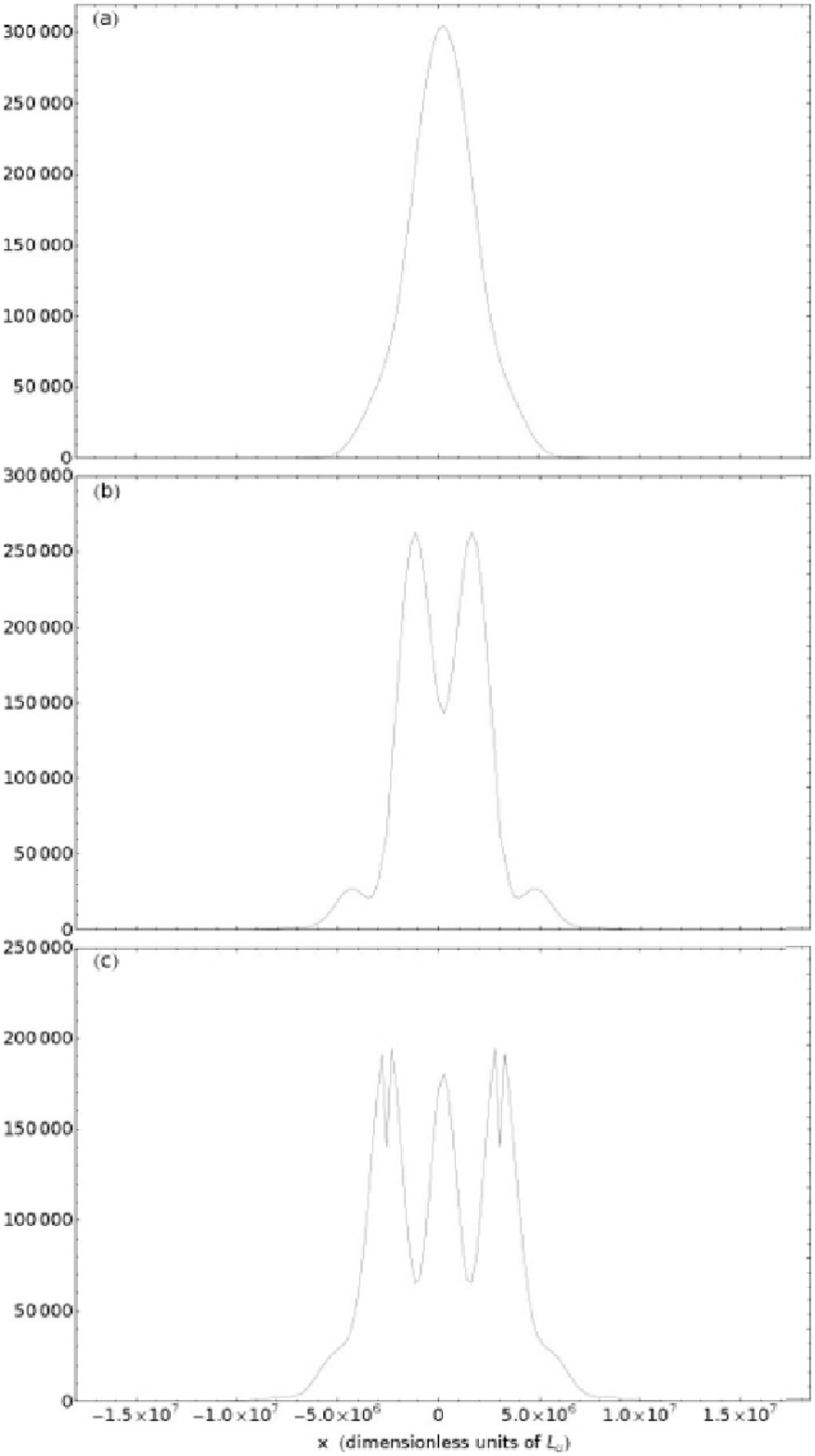,height=6.5in,width=3.25in}
\caption{Figures (a) through (c) are plots of the one-dimensional time-of-flight distributions for the double-well eigenstates $| E_1\rangle$, $| E_2\rangle$ and $| E_4\rangle$ respectively in the strongly interacting regime. The number density $n(x)$ in the ordinate is for $10^6$ double wells after a time of flight $\tau = 10^5$ (in units of $T_u$). The abscissa is shown in dimensionless units of $L_u$. }
\label{fig:tof_1lakh_tonks }
\end{figure}

\begin{figure}
\ \psfig{file=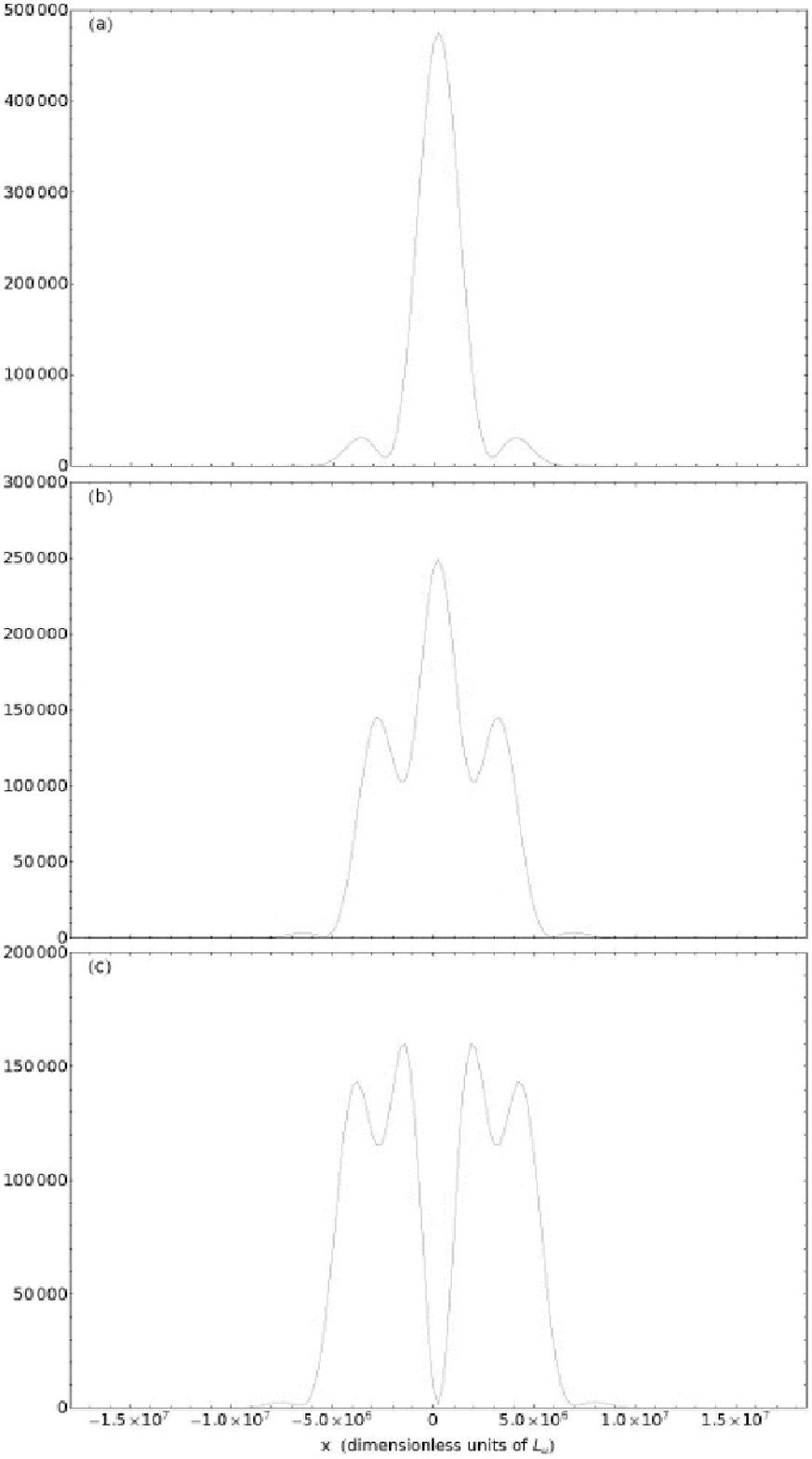,height=6.5in,width=3.25in}
\caption{Figures (a) through (c) are plots of the time-of-flight distributions for the double-well eigenstates $| E_1\rangle$, $| E_4\rangle$ and $| E_7\rangle$ respectively in the single particle regime. The number density $n(x)$ in the ordinate is for $10^6$ double wells after a time of flight $\tau = 10^5$ (in units of $T_u$). The abscissa is shown in dimensionless units of $L_u$.}
\label{fig:tof_1lakh }
\end{figure}

Numerical results for the tof distributions of the eigenstates of the double well for the strongly interacting and single particle regimes are shown in Figs~\ref{fig:tof_1lakh_tonks } and ~\ref{fig:tof_1lakh } respectively. The distributions are shown for tof $\tau=10^5$ units of $T_u$. All the dynamics is essentially independent of the characteristic length scale $L_u$ (the actual position of the well minima). For practical reasons, we choose an $L_u$ of $50$ $nm$~\cite{mypaper}. Consequently, with a $^{85}$Rb atomic mass of $85.4678$ $g·mol^{-1}$, we get a $T_u$ of about $6.7$  $\mu s$, which makes $\tau$ to be $0.67$ seconds. Using Eqn~\ref{eq:debroglie }, we get a detector distance of about $2.2$ $cm$. 

Figures~\ref{fig:tof_1lakh_tonks }(a) through (c) show the tof distributions of the states $| E_1\rangle$, $| E_2\rangle$, and $| E_4\rangle$ respectively for the strongly interacting gas detailed in section 1.
Figures~\ref{fig:tof_1lakh }(a) through (c) show the tof distributions of the states $| E_1\rangle$, $| E_4\rangle$, and $| E_7\rangle$ respectively in the single particle regime detailed in section 1. The distrubutions match the tof distributions obtained by Murphy and McCann~\cite{McCann:doublewell}, but were obtained using a finite element method that is computationally far less resource-intensive.

Time of flight fluorescence methods for profiling the wavefunction, such as measuring the momentum distribution by interrupting the particle flow with counter-propagating laser beams and then measuring fluorescence as a function of time (time of flight absorption)~\cite{fluorescense}~\cite{fluorescense:web}, will have high signal to noise ratio (compared to absorption)~\cite{raizen}.  Single shot fluorescence images should duplicate the profile shown in Figs~\ref{fig:tof_1lakh }(a)-(c) for a double well system produced by optical lattices. For a single magnetically confined double well, repeated measurements of position by the means of atom detectors, or by performing scanning tunneling microscopy on an appropriate substrate where the atoms are allowed to deposit after their tof expansion, should reproduce the required results.
\pagebreak

\section{Conclusions}
We have considered the system of two interacting bosons in a quartic double well potential of the type $V_0 (-2x^2+x^4)$. The eigenspectrum has been shown for both strongly interacting and nearly single-particle regimes, and the momentum space distributions shown for both regimes using simple numerical methods. The double well arrangement is experimentally obtainable, and the tof distributions of the lower energy states are useful information needed for doing quantum control and quantum information science problems in such systems.
\section{Acknowledgments}
The authors wish to thank the Robert A. Welch Foundation (Grant No. F-1051) for support of this work. A.R. thanks Prof. Mark Raizen for useful discussions about time-of-flight and the possibility of experiments on this system.

\pagebreak


\begin{thebibliography}{10}

\bibitem{weiman}
C.~Monroe, W.~Swann, H.~Robinson, and C.~Wieman.
\newblock Very cold trapped atoms in a vapor cell.
\newblock {\em Phys. Rev. Lett.}, 65(13):1571--1574, Sep 1990.

\bibitem{weiman:cornell}
M.H. Anderson, J.R. Ensher, M.R. Matthews, C.E. Weiman, and E.A. Cornell.
\newblock Observation of {B}ose-{E}instein {C}ondensation in a {D}ilute
  {A}tomic {V}apor.
\newblock {\em Science}, 269:198--201, July 1995.

\bibitem{ketterle}
Wolfgang Ketterle, Kendall~B. Davis, Michael~A. Joffe, Alex Martin, and
  David~E. Pritchard.
\newblock High densities of cold atoms in a dark spontaneous-force optical
  trap.
\newblock {\em Phys. Rev. Lett.}, 70(15):2253--2256, Apr 1993.

\bibitem{ketterle2}
K.~B. Davis, M.~O. Mewes, M.~R. Andrews, N.~J. {van Druten}, D.~S. Durfee,
  D.~M. Kurn, and W.~Ketterle.
\newblock {B}ose-{E}instein {C}ondensation in a gas of {S}odium atoms.
\newblock {\em Phys. Rev. Lett.}, 75(22):3969--3973, Nov 1995.

\bibitem{raizen}
C.-S. Chuu, F.~Schreck, T.~P. Meyrath, J.~L. Hanssen, G.~N. Price, and M.~G.
  Raizen.
\newblock Direct {O}bservation of {S}ub-{P}oissonian {N}umber {S}tatistics in a
  {D}egenerate {B}ose {G}as.
\newblock {\em Phys. Rev. Lett.}, 95(26):260403, Dec 2005.

\bibitem{diener}
Roberto~B. Diener, Biao Wu, Mark~G. Raizen, and Qian Niu.
\newblock Quantum tweezer for atoms.
\newblock {\em Phys. Rev. Lett.}, 89(7):070401, Jul 2002.

\bibitem{doublewell:chip}
Schumm, T., Kruger, P., Hofferberth, S., Lesanovsky, I., Wildermuth, S., Groth,
  S., Bar-Joseph, I., Andersson, L., Schmiedmayer, and J.
\newblock A {D}ouble {W}ell {I}nterferometer on an {A}tom {C}hip.
\newblock {\em Quantum Information Processing}, 5(6):537--558, December 2006.

\bibitem{doublewell:chip:nature}
T.~Schumm, S.~Hofferberth, L.~M. Andersson, S.~Wildermuth, S.~Groth,
  I.~Bar-Joseph, J.~Schmiedmayer, and P.~Kruger.
\newblock {\em Nature Physics}, 1(1):57--62, 2005.

\bibitem{Deutsch:Jessen}
Ivan~H. Deutsch and Poul~S. Jessen.
\newblock Quantum-state control in optical lattices.
\newblock {\em Phys. Rev. A}, 57(3):1972--1986, Mar 1998.

\bibitem{mypaper}
Analabha Roy and L.~E. Reichl.
\newblock Coherent {C}ontrol of {T}rapped {B}osons.
\newblock {\em Physical Review A (Atomic, Molecular, and Optical Physics)},
  77(3):033418, 2008.

\bibitem{technion:oplattice-culling}
A.~Itah, H.~Veksler, O.~Lahav, A.~Blumkin, C.~Moreno, C.~Gordon, and
  J.~Steinhauer.
\newblock Direct observation of number squeezing in an optical lattice.
\newblock Can be viewed at \url{http://arxiv.org/abs/0903.3282}, Mar 2009.

\bibitem{Monroe:Nature}
C.~Monroe, 
\newblock {\em Nature} {\bf 416}, 238 (2002).

\bibitem{Jaksch:Zoller}
D.~Jaksch, H.-J.~Briegel, J.~Cirac, C.~Gardiner, and P. Zoller, 
\newblock {\em Phys. Rev. Lett.} {\bf 82}, 1975 (1999).

\bibitem{McCann:doublewell}
D.S.~Murphy, and J.F.~McCann
\newblock {\em Phys. Rev. A } {\bf 77}, 063413 (2008).

\bibitem{metcalf:vanderstraten}
H.J. Metcalf and P.~{van der Straten}.
\newblock {\em Laser {C}ooling and {T}rapping}.
\newblock Springer Verlag, New York, 1999.

\bibitem{pethick:bec}
C.J. Pethick and H.~Smith.
\newblock {\em Bose-{E}instein {C}ondensation in {D}ilute {G}ases}.
\newblock Cambridge University Press, 2002.

\bibitem{olshanii:1d}
M.~Olshanii.
\newblock Atomic {S}cattering in the {P}resence of an {E}xternal {C}onfinement
  and a {G}as of {I}mpenetrable {B}osons.
\newblock {\em Phys. Rev. Lett.}, 81(5):938--941, Aug 1998.

\bibitem{petrov:1d}
D.~S. Petrov, G.~V. Shlyapnikov, and J.~T.~M. Walraven.
\newblock Regimes of {Q}uantum {D}egeneracy in {T}rapped 1{D} {G}ases.
\newblock {\em Phys. Rev. Lett.}, 85(18):3745--3749, Oct 2000.

\bibitem{tonks:gas}
Belen Paredes, Artur Widera, Valentin Murg, Olaf Mandel, Simon Folling, Ignacio
  Cirac, Gora~V. Shlyapnikov, Theodor~W. Hansch, and Immanuel Bloch.
\newblock {\em Nature}, 429(6989):277--281, 2004.

\bibitem{feshbach:resonance}
H.~Feshbach.
\newblock {\em Ann. Phys.}, 19(287), 1962.

\bibitem{sakurai}
J.J. Sakurai.
\newblock {\em Modern Quantum Mechanics}.
\newblock Addison-Wesley, revised edition, 1999.

\bibitem{goodman}
Joseph~W. Goodman.
\newblock {\em Introduction to {F}ourier {O}ptics}.
\newblock Roberts and Company, third edition, 2005.

\bibitem{Grondalski:etal}
John Grondalski, Paul Alsing, and Ivan Deutsch.
\newblock Spatial correlation diagnostics for atoms in optical lattices.
\newblock {\em Opt. Express}, 5(11):249--261, 1999.

\bibitem{arfken}
G~B Arfken and H~J Weber.
\newblock {\em Mathematical {M}ethods for {P}hysicists}.
\newblock Harcourt Academic Press, 5th edition, 2001.

\bibitem{fluorescense}
Paul~D. Lett, Richard~N. Watts, Christoph~I. Westbrook, William~D. Phillips,
  Phillip~L. Gould, and Harold~J. Metcalf.
\newblock Observation of {A}toms {L}aser {C}ooled below the {D}oppler {L}imit.
\newblock {\em Phys. Rev. Lett.}, 61(2):169--172, Jul 1988.

\bibitem{fluorescense:web}
Georg Raithel.
\newblock More about optical lattices.
\newblock Available as
  \url{http://cold-atoms.physics.lsa.umich.edu/projects/lattice/latticeindex.h%
tml}.

\bibitem{Dudarev:Raizen:Niu}
A.~M. Dudarev, M.~G. Raizen, and Qian Niu.
\newblock Quantum {M}any-{B}ody {C}ulling: {P}roduction of a {D}efinite
  {N}umber of {G}round-{S}tate {A}toms in a {B}ose-{E}instein {C}ondensate.
\newblock {\em Physical Review Letters}, 98(6):063001, 2007.

\bibitem{stirap:review}
Nikolay~V. Vitanov, Thomas Halfmann, Bruce~W. Shore, and Klaas Bergmann.
\newblock Laser-induced population transfer by adiabatic passage techniques.
\newblock {\em Annu. Rev. Phys. Chem.}, 52(763), 2001.

\bibitem{reichl}
L.E. Reichl.
\newblock {\em The {T}ransition to {C}haos: {C}onservative {C}lassical
  {S}ystems and {Q}uantum {M}anifestations}.
\newblock Springer {V}erlag, Berlin, 2 edition, 2004.

\end{thebibliography}
\end{document}